\documentclass[openacc]{rstransa}

\usepackage{amssymb}
\usepackage{amsmath}
\usepackage{dsfont}
\usepackage{bbm}
\usepackage[numbers,sort&compress]{natbib}

\newcommand{\nshots}{\ensuremath{s}}
\newcommand{\bra}[1]{\ensuremath{\left\langle#1\right|}}
\newcommand{\ket}[1]{\ensuremath{\left|#1\right\rangle}}

\begin{document}

\title{Towards Quantum Simulations in Particle Physics and Beyond on Noisy Intermediate-Scale Quantum Devices}

\author{
L.~Funcke$^{1,2}$, T.~Hartung$^{3,4}$, K.~Jansen$^5$, S.~K\"uhn$^4$, M.~Schneider$^{5,6}$, P.~Stornati$^{5,6}$ and X.~Wang$^{7}$}

\address{$^{1}$Center for Theoretical Physics, Co-Design Center for Quantum Advantage, and NSF AI Institute for Artificial Intelligence and Fundamental Interactions, Massachusetts Institute of Technology, 77 Massachusetts Avenue, Cambridge, MA 02139, USA\\
$^{2}$Perimeter Institute for Theoretical Physics, 31 Caroline Street North, Waterloo, ON N2L 2Y5, Canada\\
$^{3}$Department of Mathematical Sciences, University of Bath, 4 West, Claverton Down, Bath, BA2 7AY, UK\\
$^{4}$Computation-based  Science  and  Technology  Research  Center, The  Cyprus  Institute,  20  Kavafi  Street,  2121  Nicosia, Cyprus\\
$^{5}$NIC, DESY Zeuthen, Platanenallee 6, 15738 Zeuthen, Germany\\
$^{6}$Institut f\"ur Physik, Humboldt-Universit\"at zu Berlin, Zum Gro{\ss}en Windkanal 6, D-12489 Berlin, Germany\\
$^{7}$School of Physics, Peking University, 5 Yiheyuan Rd, Haidian District, Beijing 100871, China}

\subject{high energy physics, error mitigation, expressivity analysis}


\corres{Karl Jansen\\
\email{karl.jansen@desy.de}\newline\newline
\textbf{Preprint number:}\\
MIT-CTP/5325}

\begin{abstract}
    We review two algorithmic advances that bring us closer to reliable quantum simulations of model systems in high energy physics and beyond on noisy intermediate-scale quantum (NISQ) devices. The first method is the dimensional expressivity analysis of quantum circuits, which allows for constructing minimal
but maximally expressive quantum circuits. The second method is an efficient mitigation of readout errors on quantum devices. 
Both methods can lead to significant improvements in quantum simulations, e.g., 
when variational quantum\linebreak eigensolvers are used. 
\end{abstract}

\begin{fmtext}
\end{fmtext}

\maketitle

\section{Introduction}
 
The Standard Model (SM) of particle physics classifies all known elementary particles and fully describes the electromagnetic, weak,
and strong interactions. All the particles of the SM have by 
now been identified experimentally, including the infamous Higgs boson, the W- and 
Z-bosons, the photons, gluons, quarks, neutrinos, and the charged leptons with 
the electron being the most prominent example.  
The SM is extremely successful and describes physical phenomena over a large distance range from $10^{-18}$~m (the distances probed at particle colliders) to $10^{26}$~m (the size of the observable universe) without any contradiction to experiment. The predictions of the SM have been experimentally confirmed
up to 11 digits of precision, such that it can be considered the
most precisely tested theory in the history of science.

These are amazing results, in particular because there are extensive world-wide 
efforts to detect physics beyond the SM, e.g. with experiments 
such as the LHC at CERN~\cite{LHC}, Belle II at KEK~\cite{belleII}, CEBAF at JLab~\cite{jlab}, XENON1T at LNGS~\cite{XENON}, or DUNE at Fermilab
\cite{Fermilab}.
These experimental searches are strongly motivated by the fact that, despite its success, 
the SM cannot explain several experimentally observed 
physical phenomena \cite{Ellis:2012zz}. The probably most striking example is the asymmetry between matter and antimatter in our universe, which leads to our sheer existence. This asymmetry requires a substantial amount of CP-violation, which is orders of magnitude larger than predicted by the SM. At the same time, the SM predicts CP-violation for the weak and strong interactions, but no  such violation has been observed for the latter. The SM also cannot explain why the masses of the elementary particles are strongly hierarchical and why the strengths of the fundamental interactions differ by many orders of magnitude. Finally, the theoretical origins of the mysterious dark matter, dark energy, and neutrino masses remain fundamentally open questions of the SM.
In addition to these observational puzzles of the SM, there are also several theoretical issues, most importantly the missing UV-completion of quantum gravity \cite{Ellis:2012zz}. Moreover, the scalar field theory that describes the Higgs boson might become a trivial, non-interacting theory at some high-energy scale. It is currently an open question whether the Higgs self-coupling might vanish well below the Planck scale~\cite{Degrassi_2012,Bezrukov_2015}, which is the fundamental cut-off scale of the SM. If so, this could give rise to quantum triviality or vacuum stability issues, which might require extra couplings of the Higgs boson to new physics above LHC energies.

In the light of these open questions of the SM, it is of utmost importance to understand phenomena such as 
CP-violation or the matter-antimatter asymmetry from a fundamental 
point of view. To this end, we require analytical tools like perturbation theory and, crucially, numerical tools that enable us to go beyond and explore inherently non-perturbative phenomena.

The standard path to study such non-perturbative phenomena is to put the theory on a Euclidean 
space-time lattice and employ Markov Chain Monte Carlo (MCMC) methods. The lattice theory provides a non-perturbative regularization\footnote{For a
more general  
non-perturbative regularization formalism of quantum field theories using 
the $\zeta$-regularization see~\cite{hartung,hartung-phd,hartung-scott,Hartung2018,Jansen2019}.
In particular, in Refs.~\cite{hartung,Jansen2019} very first quantum simulations within 
this framework were performed.}, 
both in the ultraviolet regime through a non-zero value 
of the lattice spacing and in the infrared regime through 
a finite volume~\cite{Rothe2006,Gattringer2010}. 
Such numerical techniques have been very successful for computing various predictions of the theory behind quarks and gluons, called quantum chromodynamics (QCD). Examples are the computation of  
the hadron spectrum and their structure, fundamental parameters of QCD, and the 
order parameter of spontaneous chiral symmetry breaking~\cite{Aoki_2020}. 

However, the MCMC-based methods fail when addressing the questions of CP-violation and the matter-antimatter asymmetry. The reason is the infamous sign problem~\cite{Troyer2005}, 
which leads to complex phases that prevent the application of MCMC. 
In addition, MCMC calculations are usually performed in Euclidean space-time and therefore fail to study real-time evolution, e.g., the out-equilibrium-dynamics following heavy-ion collisions or quench dynamics. Such phenomena are important to understand, e.g., the physics of heavy-ion collisions at the LHC or the Schwinger effect leading to electron-positron production in strong electric fields.
This is exactly the point where quantum computers enter the game, which are able to efficiently simulate strongly-correlated many-body systems. In particular, quantum simulations rely on the Hamiltonian formulation in Minkowskian space-time and do not rely on MC methods, therefore circumventing the sign problem and enabling the study of real-time dynamics. Thus, quantum computing might eventually open the door to  perform simulations of the SM in three spatial dimensions, in particular in regimes that are inaccessible with other approaches.

Of course, the ambitious goal of quantum simulating the SM is presently rather far away, given that only noisy intermediate-scale quantum (NISQ) computers are currently available. Still, first quantum simulations of gauge 
theories in lower dimensions showing some of the relevant features of the SM have already been successfully performed~\cite{Martinez2016, Klco2018,Klco2019,Schweizer_2019,Yang2020,Mil_2020,atas2021su2, Ciavarella_2021,A_Rahman_2021}. 
In addition, resource efficient formulations of gauge theories
for quantum computations have been developed~\cite{Banuls2017,Celi_2020,Kaplan2020,Haase2020,Paulson2020,armon2021photonmediated}, and the Hamiltonian formulation of the CP-violating topological $\theta$-terms in three spatial dimensions
has been derived~\cite{kan2021investigating}. In general, the path towards quantum simulations of 3+1D particle physics requires many incremental steps, including algorithmic development, hardware improvement, methods for circuit design, as well as error mitigation and correction techniques. In addition, validating quantum simulation experiments is essential for obtaining reliable results, in particular in regimes which are inaccessible with MC methods. Classical Hamiltonian simulations, in particular tensor network methods, have been proven to be suitable for this task~\cite{Byrnes2002,Buyens2013, Kuehn2014,Kuehn2015,Buyens2016,Banuls2016a,Silvi2016,Buyens2017, Sala2018,Silvi2019,Banuls2018a,Funcke2019}. For detailed reviews on 
quantum simulations for high energy physics, see Refs.~\cite{Banuls2020,klco2021standard, zohar2021quantum}.

One promising way to utilize NISQ devices to efficiently perform quantum simulations are hybrid quantum-classical algorithms, most crucially the variational quantum 
eigensolver (VQE)~\cite{Peruzzo2014,McClean2016}. 
The VQE algorithm has already found numerous applications for studying benchmark models of particle physics in lower dimensions~\cite{Banuls2020,klco2021standard, zohar2021quantum}. Using VQE, the computing-intensive cost function is evaluated on the quantum computer, while 
the optimization of the circuit parameters is performed on a classical 
computer via a feedback loop.  
When performing such simulations, one faces two fundamental challenges, which we will address in this article.

First, the construction of the quantum circuit is key to compute the desired cost function. However, until recently, there had been no general scheme to guide this construction. 
We filled this gap by developing a method to custom build quantum circuits, incorporate or remove symmetries, and determine the expressivity and minimality of given 
circuits~\cite{Funcke_2021,funcke2021bestapproximation}.

Second, current quantum computers are still very noisy, which means there are large errors in the state preparation, the gate operations, and the readout process. To use quantum computers for quantum simulations, it is extremely important to mitigate, if not correct for these errors. We recently developed an efficient error mitigation method that enables us to mitigate one of the most dominant errors on current superconducting quantum hardware, which is the readout error~\cite{Funcke2020}. 

Both aspects, the dimensional expressivity analysis and the readout error mitigation, at first sight seem like technical algorithmic points, which defer from our goal to perform 
quantum simulations of high-energy physics. 
However, they are essential tools to optimize the performance of
present-day noisy quantum devices and to efficiently implement quantum simulations with current and future devices.
Thus, these tools are an important step to improve the simulations of low-dimensional benchmark models in high-energy physics and to prepare for quantum simulations in three spatial dimensions in the future.  
The following two sections will provide a detailed review of the circuit expressivity analysis and the error mitigation scheme developed by us to reach these goals.

\section{Circuit expressivity}

\begin{figure}[!h]
    \centering
    \includegraphics[width=0.9\textwidth]{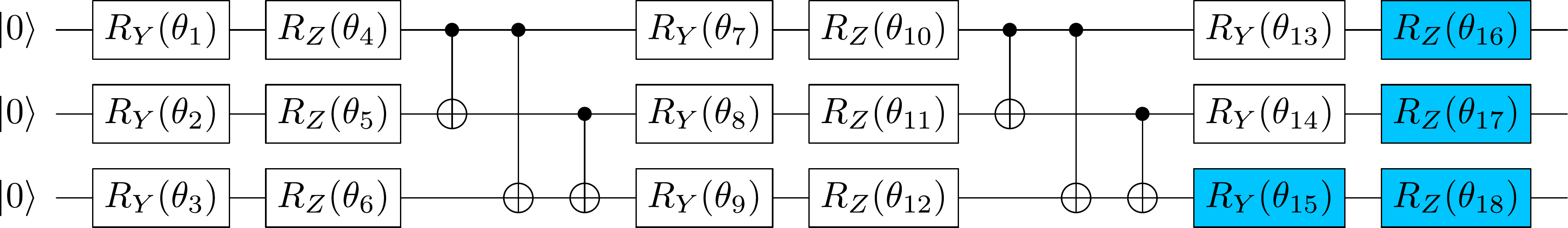}
    \caption{QISKIT's \texttt{EfficientSU2} 2-local circuit with 2 layers. The coloured 
    gates represent the gates that can be removed after the expressivity analysis
    developed here.}
\label{fig_qiskitsu2}
\end{figure}

Parametric quantum circuits (see Fig.~\ref{fig_qiskitsu2} for an example) are at the heart of VQE. It is therefore essential to find suitable ansatz circuits for obtaining the desired solution, which is usually the low-lying energy spectrum of the given problem Hamiltonian. This requires that the circuit should be equipped with sufficiently many parametric gates to express the solution. 
At the same time, the number of gates should be chosen to be minimal
in order to reduce effects of noise in a VQE simulation. 
There is no general principle for designing such minimal but maximally
expressive circuits. 
Therefore, in Refs.~\cite{Funcke_2021,funcke2021bestapproximation} 
we recently developed
a dimensional expressivity analysis, which provides a systematic and
practical way to determine whether a given quantum circuit is
sufficiently expressive and whether there are redundant parameters. In
particular, it can be used to optimize a given quantum circuit through
the removal of redundant parameters and unwanted symmetries.

\begin{figure}[!ht]
    \centering
    \includegraphics[width=2.5in]{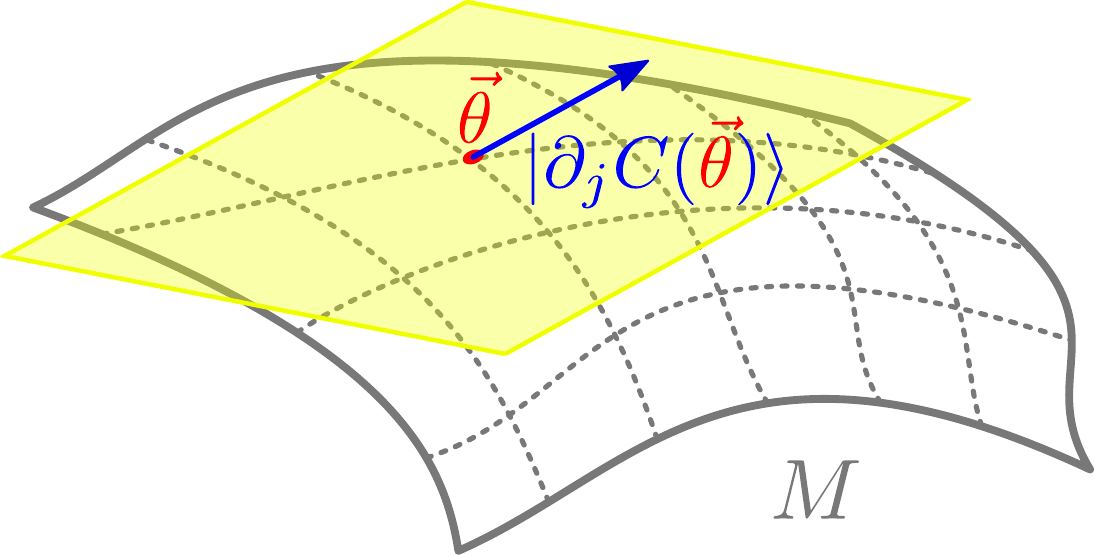}
    \caption{Representation of the tangent space (illustrated in yellow) in the point $\vec{\theta}$ (red dot) of the manifold $M$ spanned by the tangent vector $|\partial_j C(\vec{\theta})\rangle$ (blue arrow).} 
    \label{fig_tangentspace}
\end{figure}

In a quantum circuit $C(\vec{\theta})$, a number of unitary gates that depend on 
parameters $\theta$ are employed. As such, the quantum circuit $C(\vec{\theta})$ 
can be understood as a map from a parameter space into the quantum device state space. 
This leads to a manifold $M$ of states $| C(\vec{\theta})\rangle$ that 
can be reached. In addition, we have a manifold $S$ of physical states of the 
quantum device. A quantum circuit $C$ that generates only physical states therefore has a manifold $M$ that is contained in $S$.  In order to have a minimal but maximally expressive circuit, 
the co-dimension of $M$, ${\rm codim}(M) = {\rm dim}(S)-{\rm dim}(M)$, has to vanish and the number of parameters has to be equal to ${\rm dim}(S)$. 

The co-dimension can be determined by computing the tangent vectors $|\partial_j C(\vec{\theta})\rangle$ 
for a given parameter $\theta_j$, see Fig.~\ref{fig_tangentspace}, and by testing 
their linear independence. 
In particular, a parameter $\theta_{k}$ is redundant iff $|\partial_{k} C(\vec{\theta})\rangle$ 
is a linear combination of $|\partial_{j} C(\vec{\theta})\rangle, j \neq k$. 
Once the parameter $\theta_{k}$ is analyzed, the procedure can be iterated to the 
parameter $\theta_{k+1}$ until all parameters have been visited or sufficiently many independent parameters have been found to ensure the wanted expressivity. 
Having identified the set of dependent (and unnecessary) parameters, they can be removed by setting them to a suitable constant value, and 
a minimal circuit is constructed. This circuit is furthermore maximally expressive if ${\rm dim}(S)$ many independent parameters remain. Hence, the approach becomes efficiently scalable as soon as $\dim(S)$ no longer scales exponentially in the number of qubits. This is commonly seen in many physical Hamiltonians for which symmetries and entanglement restrictions imply that $\dim(S)$ only grows polynomially with the number of qubits.

For example, for QISKIT's EfficientSU2 2-local circuit for 3 qubits~\cite{Qiskit} shown 
in Fig.~\ref{fig_qiskitsu2}, such an analysis shows that the coloured 
unitary gates are redundant and can be removed. 

In practice, starting from the (real partial) Jacobian

\begin{align}
J_{k}=\left(\begin{array}{lll}\operatorname{Re}\left|\partial_{1} C\right\rangle & 
\cdots & \operatorname{Re}\left|\partial_{k} C\right\rangle \\ 
\operatorname{Im}\left|\partial_{1} C\right\rangle & \cdots & 
\operatorname{Im}\left|\partial_{k} C\right\rangle\end{array}\right),
\end{align}
we construct the matrix $S_k = J_k^{*} J_k$. The quantum circuit contains dependent parameters if ${\rm det} S_k=0$. This in turn means that the matrix $S_k$ contains vanishing eigenvalues.
The dependence of the parameters can then be determined iteratively by  computing 
the eigenvalues of $S_k$. 
Starting with $S_1$ (which is trivially independent), we move to $S_2$ and check whether it has a vanishing eigenvalue. If this is the case, then the second parameter is removed. We continue by adding one parameter at a time and checking for the smallest eigenvalue to be zero, until all parameters have been checked. A nice feature of our approach is that 
$S_k$ itself can be computed efficiently on the quantum computer 
employing one ancilla qubit, 
while the invertibility  
of $S_k$, i.e. the computation of the eigenvalues, can be performed efficiently on a classical device (see Ref.~\cite{Funcke_2021} for details). 

\subsection{Experimental results}

\begin{figure}[!h]
    \centering
    \includegraphics[width=0.8\textwidth]{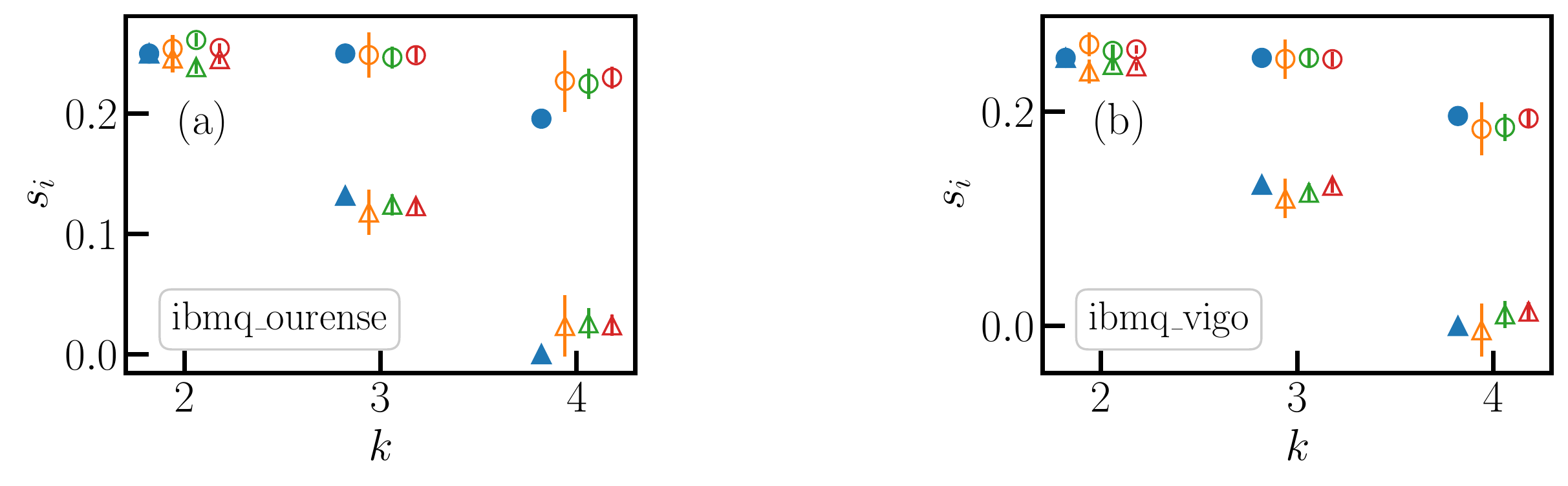}
    \caption{Single-qubit experiments on (a) ibmq\_ourense and (b) ibmq\_vigo quantum hardware. We show the smallest (triangles) and second smallest (dots) eigenvalues of the matrix $S_k$. The filled symbols are the exact solutions. The open symbols represent different statistics: orange, green and red markers stand for 1000, 4000 and 8000 shots, respectively. The error bars for the experimental data represent the uncertainty from the finite number of shots.}  
    \label{fig_sk}
\end{figure}

In Fig.~\ref{fig_sk}, we show the  
smallest and second smallest eigenvalues of $S_k$ for a 1-qubit test case. 
Our experiments have been carried out on the 
ibmq\_ourense and ibmq\_vigo quantum hardware devices. As explained in the caption 
of Fig.~\ref{fig_sk}, we have performed different numbers 
of shots, meaning different numbers of repetitions of the circuit to collect statistics of the measurement outcomes. The experimental results are shown with open symbols in Fig.~\ref{fig_sk}, while the exact solutions are shown with filled symbols. We observe that we can 
indeed identify eigenvalues that are compatible with zero within a 
given precision $\epsilon$ (see Ref.~\cite{Funcke_2021} for details). Thus, our expressivity analysis allows us to identify dependent parameters of quantum circuits on actual quantum hardware.

\subsection{Discussion} 

Through our hybrid quantum-classical approach of computing $S_k$ 
on the quantum device and testing the invertibility on a classical computer, 
we are able to escape an exponential scaling of computational 
resources, rendering our method efficient. 
As already mentioned above, in all numerical experiments we can 
identify a zero eigenvalue within a prescribed precision $\epsilon$. 
We can estimate the required computational resources 
of our method for $N_{\rm p}$ parameters in the quantum circuit: 
we need ${\cal O}(N_{\rm p}^2)$ memory, ${\cal O}(N_{\rm p}^4)$ calls on the classical machine, 
and ${\cal O}(N^2_{\rm p}/\epsilon^2)$ calls on the quantum device. Note that the 
numerical cost of analyzing a given quantum circuit is actually independent from the number of qubits employed. 

Our fundamental step of identifying redundant parameters can also be employed to remove unwanted symmetries. For example, a quantum circuit does not need to be able to generate the state $e^{i\alpha}\left|\psi\right\rangle$ if it can already generate $\left|\psi\right\rangle$. To find the parameters that only contribute such an unwanted symmetry to the circuit, we first artificially add this symmetry to the circuit. In the case of the global phase symmetry, we initialize the quantum device in $\left|0\right\rangle$ and insert a single-qubit $R_Z(\phi)$ gate into the circuit. By checking this parameter $\phi$ first, it will be considered independent. Thus, any further parameter $\theta_k$, whose independent contribution is the phase generation, will now be identified as redundant. Finally, we remove the artificially inserted $R_Z(\phi)$ gate and obtain a circuit that no longer generates arbitrary phases.

This idea can be generalized to more complicated symmetries. In general, if a continuous symmetry has dimension $d$, e.g. $d=3$ for $SU(2)$, then $d$ parameters $\phi_k$ are necessary to artificially enforce the symmetry. The symmetry can then be removed from the circuit $C(\theta)$ by extending $C$ to the artificially symmetric circuit $\tilde C(\phi,\theta)$ and testing the parameters $\phi$ before continuing to test the parameters $\theta$. 

This circuit optimization becomes even more efficient if we start with an already partially optimized circuit. For example, if the physically relevant state space is the entire quantum device state space but with certain symmetries removed, then we would like to start with a minimal but maximally expressive quantum circuit and optimize it using the removal of symmetries technique. This can be achieved using an inductive procedure. It is easy to find a minimal, maximally expressive circuit on a single qubit, e.g. $R_Y(\theta_3)R_Z(\theta_2)R_X(\theta_1)\left|0\right\rangle$.
Once we have found a minimal, maximally expressive circuit for $Q$ qubits, 
we can inductively construct a circuit for $Q+1$ qubits. This can be achieved 
by controlling an already available circuit $C_Q$ with the additionally added
qubit~\cite{funcke2021bestapproximation}. In this way, it becomes possible to construct candidates for minimal, 
maximally expressive circuits from their already existing counterparts 
with one fewer qubit. 

In practical applications, it might turn out that a maximally expressive circuit, 
although being minimal, is still too large in order to implement it on 
a given hardware device. In such cases, it is important to estimate the 
best-approximation error of the circuit that should be used. In particular, it is 
important to understand the worst-case scenario for such a situation. 
To find the best-approximation error, we have used a technique based on 
Voronoi diagrams~\cite{Voronoi1908a,Voronoi1908b}. In particular, we have determined the number 
of Voronoi points necessary to obtain a good estimate of the best-approximation
error~\cite{funcke2021bestapproximation}. This best-approximation error estimate provides an upper bound that converges to the best-approximation error for infinitely many Voronoi points. Additionally, we have provided a lower bound on the best-approximation error.
This provides a practical 
way to estimate the worst-case best-approximation error in cases where 
a minimal, maximally expressive circuit cannot be used due to the lack of 
a sufficient number of gates with high fidelity.
 
\section{Error mitigation}

Errors on quantum computers are caused by various sources of noise, including limited coherence times, imperfections in the implementations of the gate operations, as well as the measurement process. While current and future NISQ devices do not allow for quantum error correction due to the small number of qubits and their large noise levels, errors can be partly corrected through error mitigation schemes (see, e.g., Refs.~\cite{Funcke2020,Kandala:2017,Li:2017,Temme:2017,Endo:2018,Endo:2019,Kandala:2019,Tannu:2019,YeterAydeniz2019, YeterAydeniz2020, chen2021exponential,Cramer_2016}). The general idea of these schemes is to use a low-overhead procedure, e.g., to alter the circuit executed on the quantum device, to post-process the data collected from the device, to measure modified operators, or combinations thereof. In this way, the effects of quantum noise can be alleviated and more reliable estimates for expectation values of observables can be obtained. 

The final measurement can be among the most dominant sources of error, with error rates of up to $\mathcal{O}(10)$ percent~\cite{Tannu:2019}. These errors arise from bit flips, i.e.\ from erroneously reading out an outcome as $0$ given it was actually $1$, and vice versa. In the following, we will focus on a specific method developed by us~\cite{Funcke2020}, which is tailored for readout errors and can be practically implemented on existing quantum hardware. Our method scales only polynomially with the system size and is hence efficient (for a detailed comparison to previous works, see Ref.~\cite{Funcke2020}).

Throughout this section, we focus on readout errors only and neglect all other sources of error. Thus, we assume that the quantum device prepares a pure state $\ket{\psi}$, which we measure in the computational basis. In order to be able to obtain the expectation value of an observable, we have to run the circuit a number of times and collect statistics of the measurement outcomes. Just as in the previous section, we refer to this number of repetitions as the number of shots $\nshots$. Moreover, we assume that readout errors caused by bit flips between different qubits are uncorrelated.

For an illustration of the basic idea, let us consider a very simple Hamiltonian that consists only of the operator $Z$, i.e. the third Pauli matrix. We want to measure the expectation value of this Hamiltonian in an arbitrary pure single-qubit state, $|\psi\rangle=c_{1}|0\rangle+c_{2}|1\rangle$, where the complex coefficients fulfill the normalization condition $|c_1|^2+|c_2|^2=1$. The exact energy then evaluates to
\begin{equation}
        E_{Z}=\left\langle \psi\left|Z\right| \psi\right\rangle=\left\langle 0\left|c_{1}^{*} Z c_{1}\right| 0\right\rangle +\left\langle 1\left|c_{2}^{*} Z {c_{2}}\right| 1\right\rangle = \left|c_1\right|^2 - \left|c_2\right|^2\; .
    \label{eq:exactE} 
\end{equation}
This noise-free result changes in the presence of bit-flip errors in the readout process, i.e., due to the misidentification of the outcome 0 as 1 and vice versa. For illustrative purposes, let us first assume that the probabilities for misidentifying the outcome 0 as 1 and the outcome 1 as 0 are the same. For this case, we give in Tab.~\ref{tab:bitflip} the possibilities for correct and {\em wrong} measurements.
\begin{table}
    \centering
    \begin{tabular}{llll}
    \hline \hline Outcome & Measured Energy &  Operator & Probability \\
    \hline No bit flips               & $E_{Z}=+\left|c_{1}\right|^{2}-\left|c_{2}\right|^{2}$        & $\phantom{-}Z$           & $(1-p)^{2}$ \\
    $0 \rightarrow 1,1 \rightarrow 1$ & $E_{1}=-\left|c_{1}\right|^{2}-\left|c_{2}\right|^{2}$        & $-\mathds{1}$            & $p(1-p)$ \\
    $0 \rightarrow 0,1 \rightarrow 0$ & $E_{2}=+\left|c_{1}\right|^{2}+\left|c_{2}\right|^{2}=-E_{1}$ & $\phantom{-}\mathds{1}$  & $(1-p) p$ \\
    $0 \rightarrow 1,1 \rightarrow 0$ & $E_{3}=-\left|c_{1}\right|^{2}+\left|c_{2}\right|^{2}=-E_{Z}$ & $-Z$                     & $p^{2}$ \\
    \hline \hline
    \end{tabular}
    \caption{List of possible outcomes of measuring the energy of the operator $Z$ in a state $|\psi\rangle=c_{1}|0\rangle+c_{2}|1\rangle$ and their probabilities of occurrence for different cases of bit-flips.}
    \label{tab:bitflip}
\end{table}
From Tab.~\ref{tab:bitflip} we see that instead of the exact result $E_Z$, we instead obtain an \emph{expected value of the observable energy to bit flips} $\tilde{E}_{Z}$, which evaluates to
\begin{equation}
    \tilde{E}_{Z}=(1-p)^{2} E_{Z}+ p(1-p)\left(E_{1}+E_{2}\right)+p^{2} E_{3}=(1-2 p) E_{Z}\; .
    \label{eq:noisyE}
\end{equation} 
From Eq.~\eqref{eq:noisyE} it becomes apparent that one can reconstruct the exact energy from the noisy result, as long as one knows the bit-flip probability and $p\neq 1/2$.

Alternatively, instead of treating the bit-flips as part of the measurement process, we can change the point of view and consider them as part of the operator to be measured. Thus, we consider random operators to be measured. In Tab.~\ref{tab:bitflip}, we also list the corresponding operators for our single-qubit example. The expected value of the noisy operator subject to bit flips is then given by
\begin{equation}
    \mathbb{E}\tilde{Z}=(1-p)^{2} Z+ p(1-p)\left(-\mathds{1}+\mathds{1}\right)-p^{2} Z=(1-2 p) Z\;.
    \label{eq:noisyOp}
\end{equation}
This change in point of view also allows to generalize the method to arbitrary operators acting on more than a single qubit. In the following, we illustrate this with a two-qubit example and now consider the general case of \textit{arbitrary} bit-flip probabilities for the different qubits. Performing a similar analysis as for the single-qubit case above, and inverting the corresponding equation, we obtain 
\begin{align}
    \begin{aligned} 
        Z_{2} \otimes Z_{1}=& 
        \frac{1}{\gamma\left(Z_{2}\right) \gamma\left(Z_{1}\right)} \mathbb{E}\left(\tilde{Z}_{2} \otimes \tilde{Z}_{1}\right) 
        -\frac{\gamma\left(\mathbbm{1}_{1}\right)}{\gamma\left(Z_{2}\right) \gamma\left(Z_{1}\right)} \mathbb{E}\left(\tilde{Z}_{2}\right) \otimes \mathbbm{1}_{1} \\ &
        -\frac{\gamma\left(\mathbbm{1}_{2}\right)}{\gamma\left(Z_{2}\right) \gamma\left(Z_{1}\right)} \mathbbm{1}_{2} \otimes \mathbb{E}\left(\tilde{Z}_{1}\right) 
        +\frac{\gamma\left(\mathbbm{1}_{2}\right) \gamma\left(\mathbbm{1}_{1}\right)}{\gamma\left(Z_{2}\right) \gamma\left(Z_{1}\right)} \mathbbm{1}_{2} \otimes \mathbbm{1}_{1}\; . 
    \end{aligned}
    \label{eq:inverted}
\end{align}
In the expression above the factors $\gamma\left(O_{q}\right)$ are given by  
\begin{align}
    \gamma\left(O_{q}\right)&:= \left\{\begin{array}{ll}1-p_{q, 0}-p_{q, 1} & \text { for } O_{q}=Z_{q} \\ 
    p_{q, 1}-p_{q, 0} & \text { for } O_{q}=\mathbbm{1}_{q} .\end{array}\right\}\; ,
    \label{eq:gammas}
\end{align}
where $p_{q,0}$ denotes the probability that a bit-flip from 0 to 1 occurs on qubit $q$ and $p_{q,1}$ denotes the probability for a bit-flip from 1 to 0 on qubit $q$. Just as before, $\mathbb{E}$ denotes the expectation value of the noisy operator subject to bit flips. 

From the considerations above, we see that Eq.~\eqref{eq:inverted}
allows us to obtain the correct expectation value of a two-qubit operator by measuring noise-afflicted expectation values of $Z_1$, $Z_2$ and $Z_1\otimes Z_2$ on the quantum device and combining them with factors that only depend on the known bit-flip probabilities. 
A key observation is that the statistical expectation value $\mathbb{E}\left(\tilde{Z}_{Q} \ldots \tilde{Z}_{1}\right)$ of $Q$ qubits can be factorized in single-qubit expectation values 
\begin{align}
    \mathbb{E}\left(\tilde{Z}_{Q} \ldots \tilde{Z}_{1}\right)=\mathbb{E} \tilde{Z}_{Q} \cdots \mathbb{E} \tilde{Z}_{1}\; .
    \label{eq:factorize}
\end{align}
Note that $\mathbb{E}$ denotes the expectation value of the noisy operator subject to bit flips, which should not be confused with the quantum mechanical expectation value
of the operator. An inductive proof of Eq.~\eqref{eq:factorize} can be found in Ref.~\cite{Funcke2020}. This equation is the reason why the method of readout error mitigation developed in Ref.~\cite{Funcke2020} only scales polynomially and hence is efficient for $k$-local Hamiltonians, i.e., for most model systems in high energy and condensed matter physics.

\subsection{Experimental results} 

\begin{figure}[!h]
    \centering    
    \includegraphics[width=0.4\textwidth]{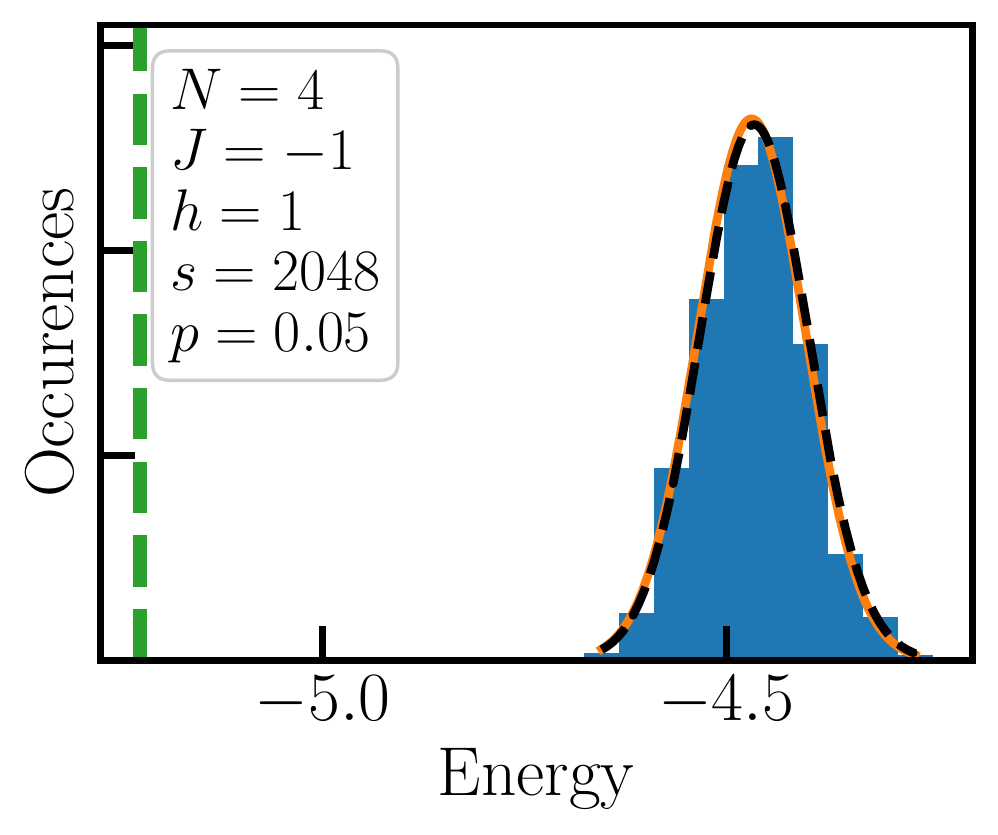}    
    \caption{Energy histogram for the transverse field Ising model for $2048$ experiments, $N=4$ qubits, and couplings $J=-1$, $h=1$. For each of the experiments, we used $\nshots=2048$ shots and a homogeneous bit-flip probability $p_{q,b} \equiv p=0.05$ for all qubits.  The vertical dashed green line indicates the true ground-state energy, the solid orange line the theoretical prediction for the noisy measurement outcomes, and the dashed black line a fit to the data.}
    \label{fig:ti}
\end{figure}

As a first non-trivial example to test our readout error mitigation scheme, we have considered the transverse field Ising (TI) model 
\begin{equation}\label{ti}
    \mathcal{H}_{\rm TI}=J\sum_{i=1}^L Z_iZ_{i+1}+h\sum_{i=1}^L X_i\; .
\end{equation} 
Here, $J$ and $h$ are the nearest neighbor and magnetic field coupling, respectively. Note that the model shows a quantum phase transition when $\left|J/h\right| = 1$, and can hence be considered as a good benchmark model for studying critical phenomena. We have performed numerical computations simulating a noise readout process with bit flips, and measured the resulting ground-state expectation value of the TI Hamiltonian for various parameters and different numbers of shots $\nshots$. The obtained histogram for one example with couplings $J=-1$, $h=1$ and four qubits is shown in Fig.~\ref{fig:ti}. There, the vertical dashed line is the true ground-state energy, and it can be seen that the mean of the measured histogram is clearly shifted away from the exact result due to the bit flips. However, our method allows to predict the form of the histogram, and the orange line in the figure represents the prediction, which basically coincides with a fit to the histogram indicated by the black solid line. This demonstrates that our method is able to reproduce the measured energy histogram, which in turn allows to reconstruct the correct ground-state energy.   

\begin{figure}[!htp]
    \centering
    \includegraphics[width=0.95\textwidth]{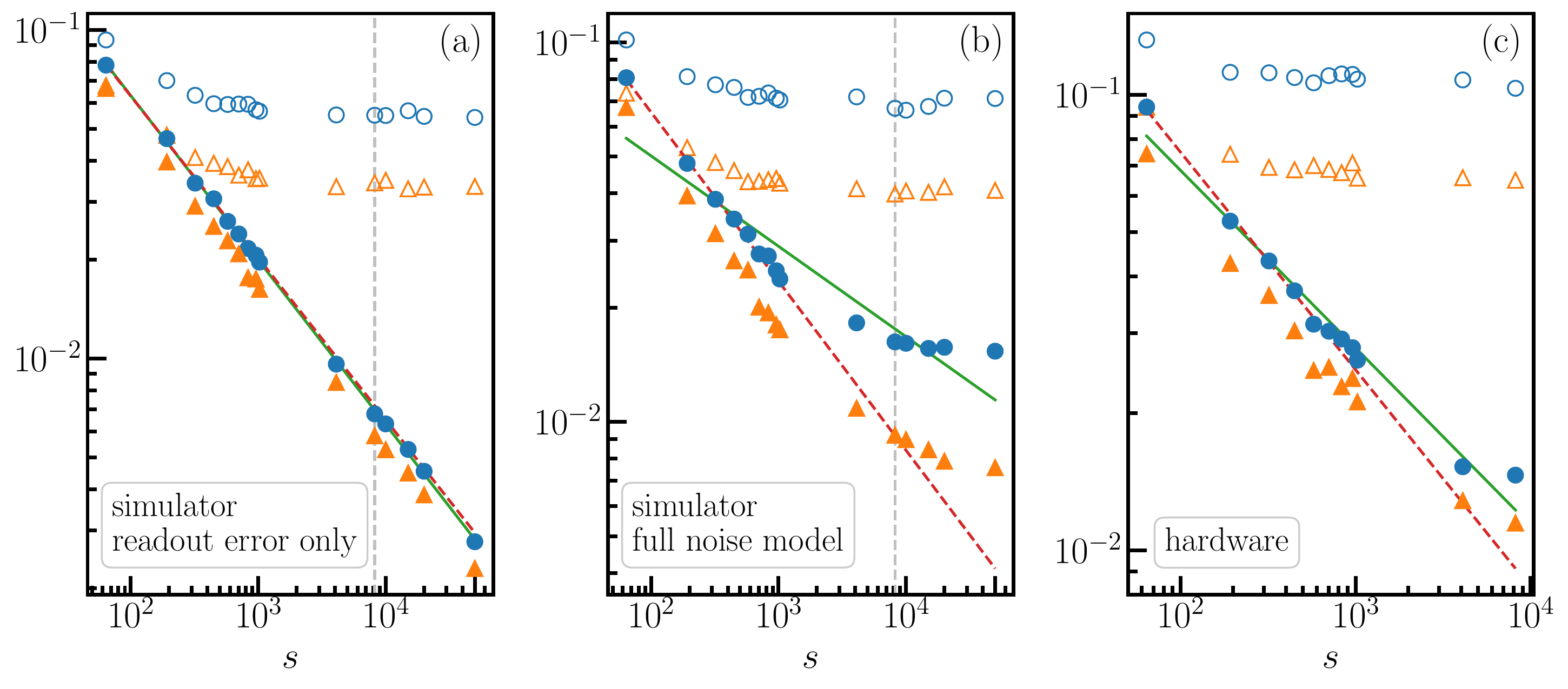}
    \caption{Mean value (blue dots) and standard deviation (orange triangles) for the 
absolute error of $\langle \Psi | Z_2\otimes Z_1 | \Psi\rangle$ 
after applying the correction procedure (filled symbols) and without it (open symbols) as a function of the number of shots $\nshots$. Each data point is obtained by carrying out 1024 experiments. 
The different panels correspond to (a) a classical simulation of two qubits of the ibmq\_london quantum device taking into account readout noise only, (b) the same simulation with a noise model emulating the full hardware noise, and (c) data obtained from the actual quantum hardware. The lines represent a power-law fit to the mean error, where the green line corresponds to a fit to all of our data points, and the red dashed lines corresponds to a fit including only the four smallest values of $\nshots$. The vertical gray dashed lines in panels (a) and (b) indicate the maximum number of shots that can be executed on the actual quantum hardware.}
    \label{fig:data_ibmq_london}
\end{figure}

An important question is the scaling of the error of the measured ground-state energy as a function of shots $s$. To this end, we have performed simulations for two qubits using Qiskit~\cite{Qiskit}, and compared the mean value and the standard deviation of the error
\begin{equation}
    \left| \bra{\psi}\tilde{Z}_2\otimes\tilde{Z}_1\ket{\psi}_\text{measured} - \bra{\psi}Z_2\otimes Z_1\ket{\psi}_\text{exact} \right|
\end{equation}
obtained for 1024 randomly drawn wave functions $\ket{\psi}$. In Fig.~\ref{fig:data_ibmq_london}, we show the scaling of the mean value of the error as well as the standard deviation as a function of the number of shots, both with and without mitigation. Panel (a) contains the data from a simulation with readout error only, and panel (b) the data obtained using a noise model that mimics the full noise of the quantum hardware. The data are fitted to a power law $as^{-\beta}$ with the green line taking all data into account, whereas the red line only fits the lowest four data points. 

As can be seen, for the case of readout noise only (cf.\ Fig.~\ref{fig:data_ibmq_london}(a)) and our error mitigation method applied, the expected error scaling of $\propto s^{-\beta}$ is observed with $\beta$ compatible with $0.5$, see Tab.~\ref{tab:exponents}. When the full noise model is switched on (see Fig.~\ref{fig:data_ibmq_london}(b)), deviations from this scaling at a number of shots $\mathcal{O}(10^3)$ are found. At this level, the readout error is almost fully corrected and other sources of errors become dominant, which require different noise mitigation techniques. Thus, we find a large reduction of the readout error compared to the results without any mitigation. 

In Fig.~\ref{fig:data_ibmq_london}(c) we show the results of the mean error and the standard deviation for running our experiments on the actual ibmq\_london quantum hardware. Also here we observe an order of magnitude improvement in the errors and a  scaling in the number of shots compatible with $\propto s^{-\beta}$, see Tab.~\ref{tab:exponents} for the values of $\beta$. Figure~\ref{fig:data_ibmq_london}(c) demonstrates that our error mitigation approach is working in practice and can hence be very useful in future applications for performing quantum simulations. 

\begin{table}
    \centering
    \begin{tabular}{llll}
        \hline \hline  & Simulation      & Simulation   & Hardware \\
                       & (readout only)  & (full noise) & \\
        \hline 
        full dataset         & 0.501  & 0.238 & 0.390  \\
        4 lowest data points & 0.492  & 0.446 & 0.478 \\         
        \hline \hline
    \end{tabular}
    \caption{Exponents $\beta$ of power-law fits $a\nshots^{-\beta}$ to our simulated and hardware data on the ibmq\_london quantum device.}
    \label{tab:exponents}
\end{table}

\begin{figure}[!h]
    \centering    
    \includegraphics[width=0.95\textwidth]{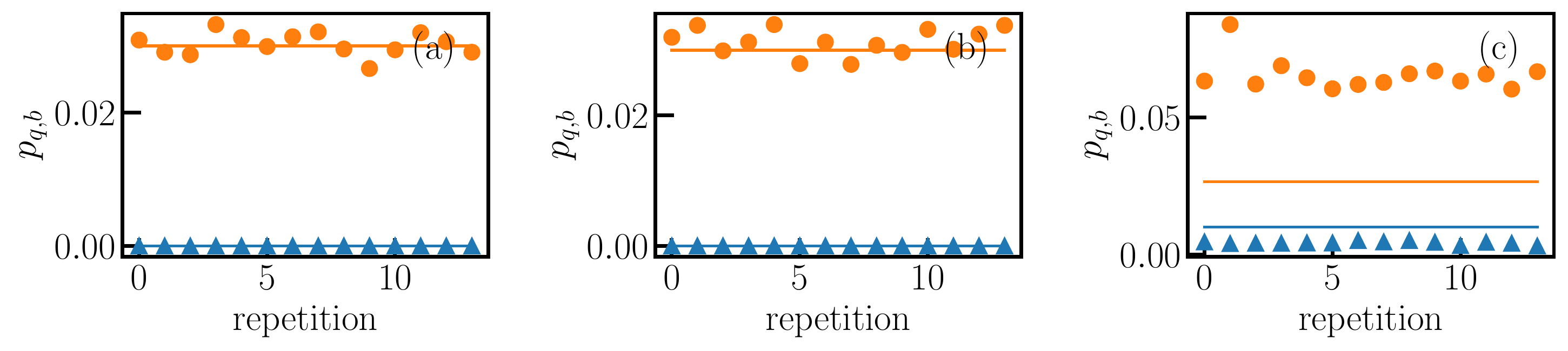}
    \caption{Bit-flip probabilities $p_{0,0}$ (blue triangles) and $p_{0,1}$ (orange dots) for the single qubit case as a function of repetitions of the experiment using (a) a noise model mimicking the readout errors of of ibmq\_london only, (b) a noise model mimicking the full the full noise of ibmq\_london, and (c) the actual quantum hardware. The solid lines correspond to the data provided by the noise model.}
    \label{fig:drift}
\end{figure}

A key ingredient of our approach is the calibration of the readout errors of the qubits used in our experiments. Such calibrations have to be carried out before each run since there is a strong drift in the bit-flip probabilities. This is shown in Fig.~\ref{fig:drift}. While the bit-flip probability is consistent with the values of the noise model in the simulator runs (see Figs.~\ref{fig:drift}(a) and~\ref{fig:drift}(b)), the values observed on real hardware differ significantly from those provided in the noise model (cf.~Fig.~\ref{fig:drift}(c)). This requires a calibration of the bit-flip probabilities before a quantum simulation is carried out in order to apply our error mitigation scheme.  
 
\subsection{Discussion} 

On the superconducting devices used in our experiments, we generally observe that readout errors of different qubits are uncorrelated to a good approximation. Thus, we have neglected multi-qubit correlations in this work. However, such correlations can be taken into account as discussed in Ref.~\cite{Funcke2020} through 
multi-qubit calibrations.

Our method can also be applied, at least in principle, to other sources of noise. If, for example, we consider the qubit relaxation error with characteristic time scale $T_1$ (decay of $|1\rangle$ to $|0\rangle$)~\cite{Qiskit}, the measurement outcome for the $Z$ operator corresponds to $\tilde{Z} = p(t)Z + [1-p(t)]\mathbbm{1}$, where $p(t)=\exp\left[-(t/T_1)\right]$ is the probability that $|1\rangle$ has not yet decayed. Then the exact operator $Z$ is obtained from the noisy one by $Z = [1/p(t)]\tilde{Z} - \{[1-p(t)]/p(t)\}\mathbbm{1}$, meaning that we can again reconstruct the operator $Z$ through noisy measurements. 

Another advantage of our method is that it can be pre-processed, meaning the operator to be measured can be directly replaced with an appropriate linear combination of operators whose quantum mechanical expectation value subject to bit flips corresponds to the true expectation value of the original operator. This allows for easy use of our method with existing software frameworks. For example, many libraries provide functions implementing the VQE, which typically ask for a Hamiltonian as an input parameter. Passing the appropriate bit-flip corrected Hamiltonian as an argument, the readout errors can be mitigated without additional effort.

As already mentioned above, our error mitigation method scales polynomially for $k$-nearest neighbor Hamiltonians and adds a moderate overhead for non-local Hamiltonians. Since most models in high energy or condensed matter physics have Hamiltonians with nearest or next-to-nearest interactions, our error mitigation scheme is therefore efficient.

\section{Conclusion}

In this article, we have reviewed two algorithmic 
advances towards eventually simulating models in high-energy physics 
and beyond
with quantum devices. 
The motivation for such quantum simulations is twofold: first, there are several experimentally observed phenomena, such as the large amount of CP-violation in the universe, which cannot be explained by the SM; and second, several of these phenomena cannot be simulated with classical MCMC methods. 

When realizing simple benchmark models on a quantum computer, for example the 1+1 dimensional 
Schwinger model, the results can be erroneous even for a small 
number of qubits. The reasons are various 
sources of noise on present quantum devices. 
It is therefore highly desirable to develop methods to reduce the number of noisy quantum gates and to mitigate the noise before embarking on simulations of realistic models. 

To this end, we first developed the dimensional expressivity analysis, which allows us to identify redundant gates in a given quantum circuit. 
This in turn leads to the construction of minimal, but maximally 
expressive quantum circuits.  
In this way, we reduced the number of noisy gates and ensured that the constructed quantum circuit can reach the desired 
manifold of quantum states. If a maximally expressive quantum 
circuit cannot be realized, 
we developed an additional technique based on Voronoi diagrams to bound the 
best-approximation error, which helps us to estimate how far away the quantum computation can be from the 
desired state space.

As a second step, we developed a very general and efficient scheme for 
mitigating the readout error, which is presently among the dominating 
error on superconducting quantum computers. Our scheme relies on the readout 
error calibration of the used qubits and allows to evaluate the 
exact Hamiltonian from noisy measurements. We demonstrated the performance of our method by 
quantum computations both on the IBM-Q simulator and the IBM-Q hardware. Our algorithm shows a polynomial 
scaling in the number of qubits and is therefore efficient, at least when $k$-nearest neighbour interactions are considered. 
We are presently working on extending this error mitigation 
scheme to other sources of noise on quantum computers. 

These advances in the circuit expressivity analysis and error mitigation tremendously 
help to render quantum computer simulations more reliable. Thus, one can 
eventually come back
to the targeted simulations of high-energy physics models, with the goal of addressing 
at least some of the 
remaining open questions of the SM of particle physics.

\enlargethispage{20pt}

\funding{Research at Perimeter Institute is supported in part by the Government of Canada through the Department of Innovation, Science and Industry Canada and by the Province of Ontario through the Ministry of Colleges and Universities. L.F.\ is partially supported by the Co-Design Center for Quantum Advantage (C2QA) under subcontract number 390034, by the DOE QuantiSED Consortium under subcontract number 675352, by the National Science Foundation under Cooperative Agreement PHY-2019786 (The NSF AI Institute for Artificial Intelligence and Fundamental Interactions, http://iaifi.org/), and by the U.S.\ Department of Energy, Office of Science, Office of Nuclear Physics under grant contract numbers DE-SC0011090 and DE-SC0021006. S.K.\ acknowledges financial support from the Cyprus Research and Innovation Foundation under project ``Future-proofing Scientific Applications for the Supercomputers of Tomorrow(FAST)'', contract no.\ COMPLEMENTARY/0916/0048.}

\ack{We thank Constantia Alexandrou, Giovanni Ianelli, Georgios Polykratis, Davide Racco, and Tom Weber for many useful discussions.} 


\bibliographystyle{unsrtnat}
\bibliography{Papers}

\end{document}